# Systematic Literature Review: Explainable AI Definitions and Challenges in Education

*Literature Review*

**Zaid M. Altukhi**
University of Technology Sydney[1],
Islamic University of Madinah[2]
[1]Sydney, Australia, [2]Madinah, KSA
Zaid.m.altukhi@student.uts.edu.au

**Dr Sojen Pradhan**
University of Technology Sydney
Sydney, Australia
sojen.pradhan@uts.edu.au

## Abstract

*Explainable AI (XAI) seeks to transform black-box algorithmic processes into transparent ones, enhancing trust in AI applications across various sectors such as education. This review aims to examine the various definitions of XAI within the literature and explore the challenges of XAI in education. Our goal is to shed light on how XAI can contribute to enhancing the educational field. This systematic review, utilising the PRISMA method for rigorous and transparent research, identified 19 relevant studies. Our findings reveal 15 definitions and 62 challenges. These challenges are categorised using thematic analysis into seven groups: explainability, ethical, technical, human-computer interaction (HCI), trustworthiness, policy and guideline, and others, thereby deepening our understanding of the implications of XAI in education. Our analysis highlights the absence of standardised definitions for XAI, leading to confusion, especially because definitions concerning ethics, trustworthiness, technicalities, and explainability tend to overlap and vary.*

## Introduction

Artificial Intelligence (AI) is transforming many tasks across various sectors, including finance, healthcare, and education. Although AI-driven systems have grown increasingly sophisticated and effective in making decisions (Kaur et al., 2022), the results produced by them are difficult to trust because of their opaque, black-box characteristics (Ali et al., 2023; Rachha & Seyam, 2023). There are ethical considerations and concerns about algorithmic fairness that need to be addressed to ensure that AI decisions do not have a bias or discriminate against certain groups of users.

Explainable AI (XAI) has emerged as a crucial development for overcoming these challenges. XAI aims to uncover the vagueness of AI's decision-making processes by offering insights (clarity) into how AI systems reach their conclusions (Galhotra et al., 2021; Rai, 2020). The process of making AI more transparent seeks to build trust, facilitate a deeper understanding of AI operations and improve confidence in these systems (Hanif et al., 2021; Hasib et al., 2022). In the context of education, it is important to ensure that AI-driven systems maintain high standards to protect end-user's privacy and offer the decisions and outcomes expected in order to adapt and trust the results of these systems.

This review selects the recent research studies on XAI in education using Preferred Reporting Items for Systematic Reviews and Meta-Analyses (PRISMA) methodology, as well as identifies the definitions and major challenges within the field. The methodology section shows the eligibility criteria, and the selection process used to select the papers that have been used in this review. In addition, the result section shows the findings from the selected papers, with the different definitions of XAI retrieved from those papers and the challenges mentioned. We also used thematic analysis to classify the identified challenges into several categories. The discussion section presents the connection between each category of the challenges. Finally, we conclude this review with some suggestions for the research agenda for the future.

# Methodology

The procedure for identifying relevant scientific literature adhered to the PRISMA guidelines for systematic literature reviews, as recommended by Page et al. (2021). This approach was customised to filter the eligibility criteria, search strategy, and paper selection process required for this topic. Initially, we embarked on two key tasks: delineating the various definitions and identifying the challenges linked to XAI. The review's eligibility criteria were strictly applied, specifying the inclusion and exclusion of publications. We also used the thematic analysis proposed by (Braun & Clarke, 2012) to categorise the challenges mentioned in the chosen articles. The thematic approach ensures that the classification builds on a scientific basis.

## *Eligibility Criteria*

This study employed specific inclusion and exclusion criteria to ensure the quality and relevance of the literature. Included in the publications were journal articles and conference papers published in English between 2020 and August 2023. Eligible study designs encompassed primary research using quantitative, qualitative, mixed-methods, case study, or experimental designs that specifically investigated the application of XAI within an educational context. Publications were excluded if they were books, book chapters, notes, ebooks, video conferences, theses, non-peer-reviewed publications, or any other irrelevant works. Additionally, any publication not in English or lacking a direct focus on the intersection of XAI or education was omitted from this review.

## *Search Strategy*

A systematic search strategy was implemented to identify relevant publications across three major academic databases: IEEE, Scopus, and ACM. The search spanned publications from 1 January 2020 to August 2023, when this research was conducted. Specific search terms were employed, including combinations of explainable AI, XAI, education pedagogy, learning, teaching, and assessment.

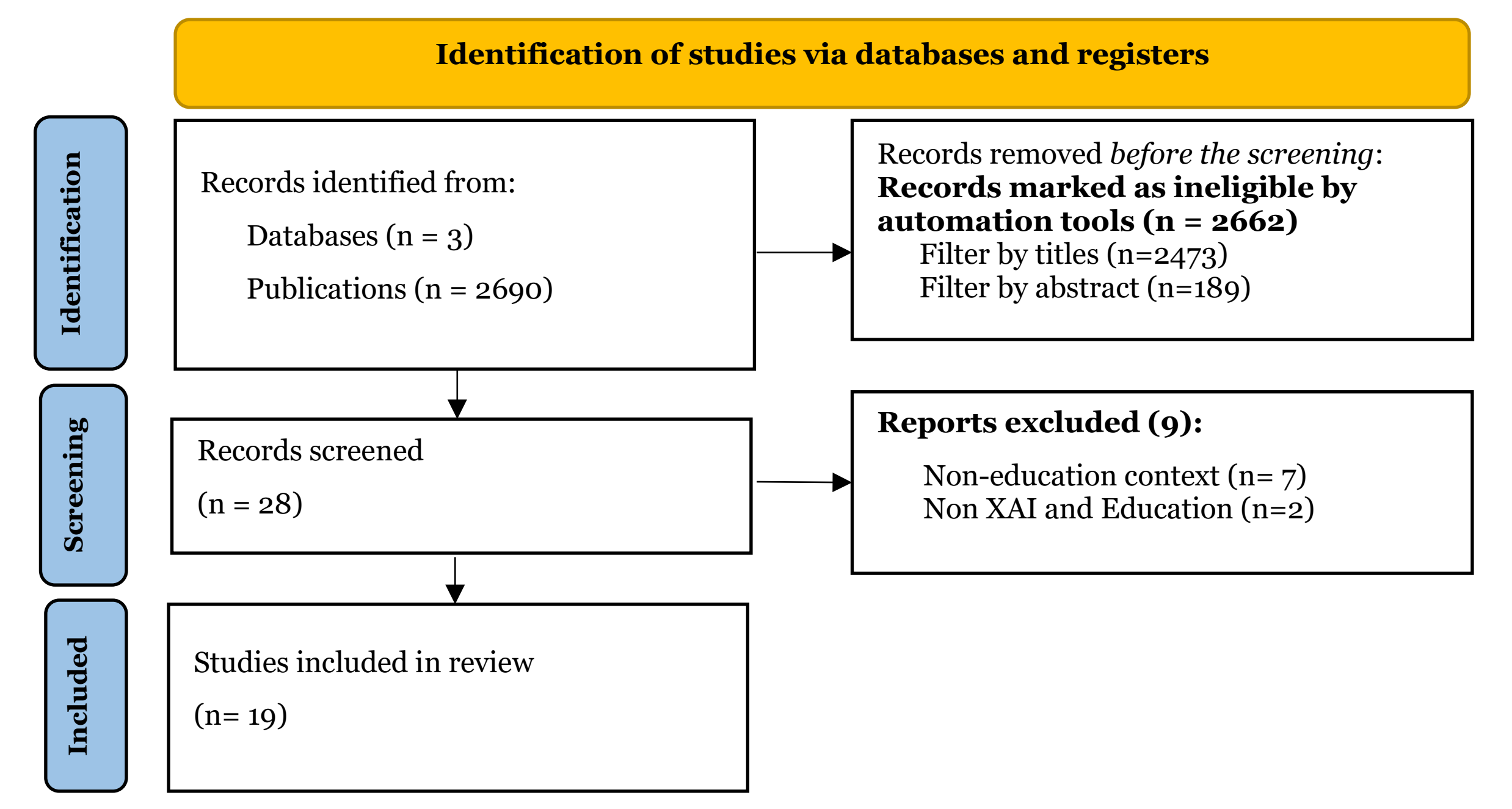

**Figure 1. PRISMA methodology for the systematic review.**

### *Selection Process*

The selection process for the systematic review was organised and executed in several phases to ensure the precision and relevance of the selected studies. Initially, the process began with an extensive screening of titles and abstracts from a preliminary retrieval of 2,690 studies that potentially met this review's objectives without delving into the full texts.

Subsequently, a Python script was utilised to enhance the search, specifically targeting the titles containing XAI and education. This script filtered out publications by identifying titles containing key terms such as 'Explainable' or 'XAI,' which resulted in the exclusion of 2,473 titles, thereby refining the selection of relevant studies.

The filtering process then proceeded to a thorough review of the 236 remaining abstracts. The objective was to confirm that the abstracts contained the designated keywords (XAI, Explainable) and incorporated terms pertinent to the educational context. This stage was critical to ensure that the selection of studies matched the review's thematic emphasis on the implementation of XAI in educational environments, leading to the exclusion of 189 articles.

Next, in the screening stage, all full text of the remaining 28 articles were read and examined. During this process, a number of studies were excluded based on the aforementioned criteria. Specifically, seven studies were excluded for not being related to education and two for lacking relevance to XAI and education. This exclusion phase was instrumental in ensuring that only studies with direct relevance to the research were considered.

Finally, full-text copies of 19 articles were obtained for an in-depth evaluation, representing a comprehensive and systematic selection process.

### *Categorisation Process*

In line with the research objective, we categorised challenges associated with XAI within 19 articles. This process is divided into six phases as outlined by (Braun & Clarke, 2006, 2012): 1) familiarisation with the data, 2) generating initial codes, 3) generating themes, 4) reviewing potential themes, 5) naming themes, and 6) producing the report.

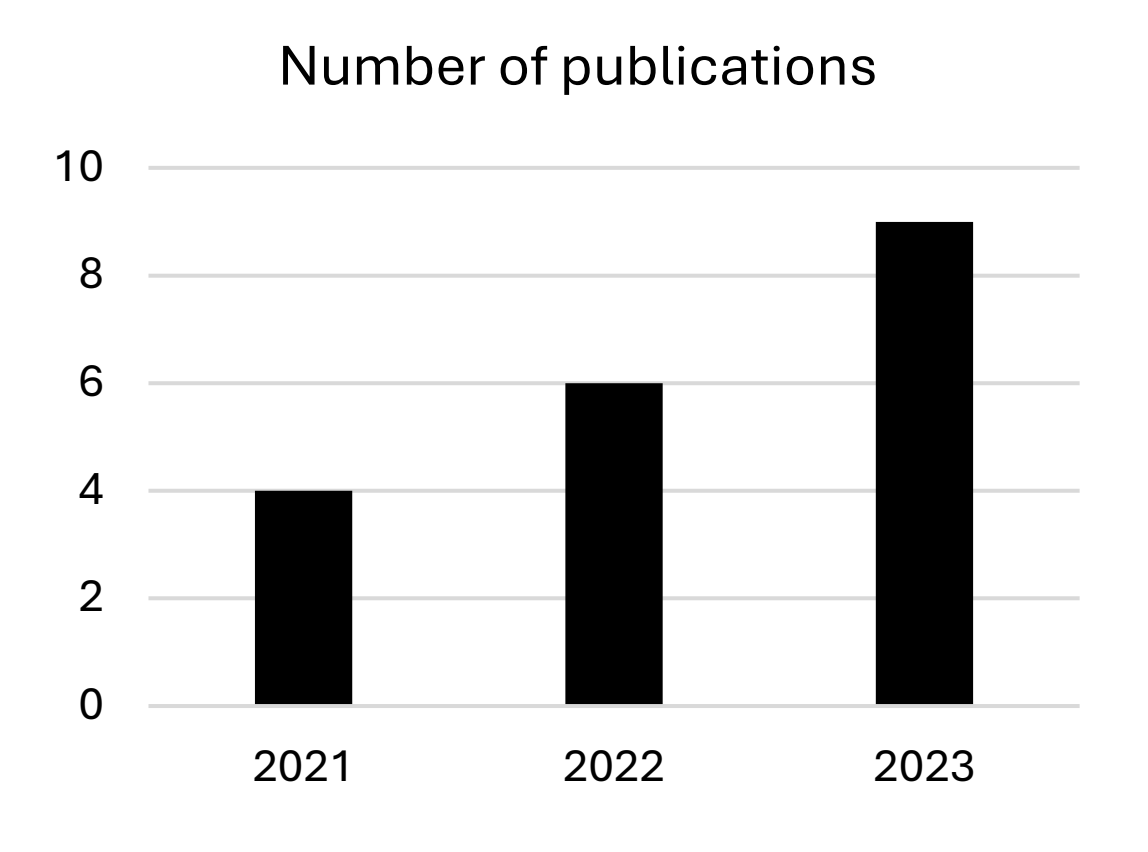

**Figure 2. Year-wise analysis of the reviewed publications.**

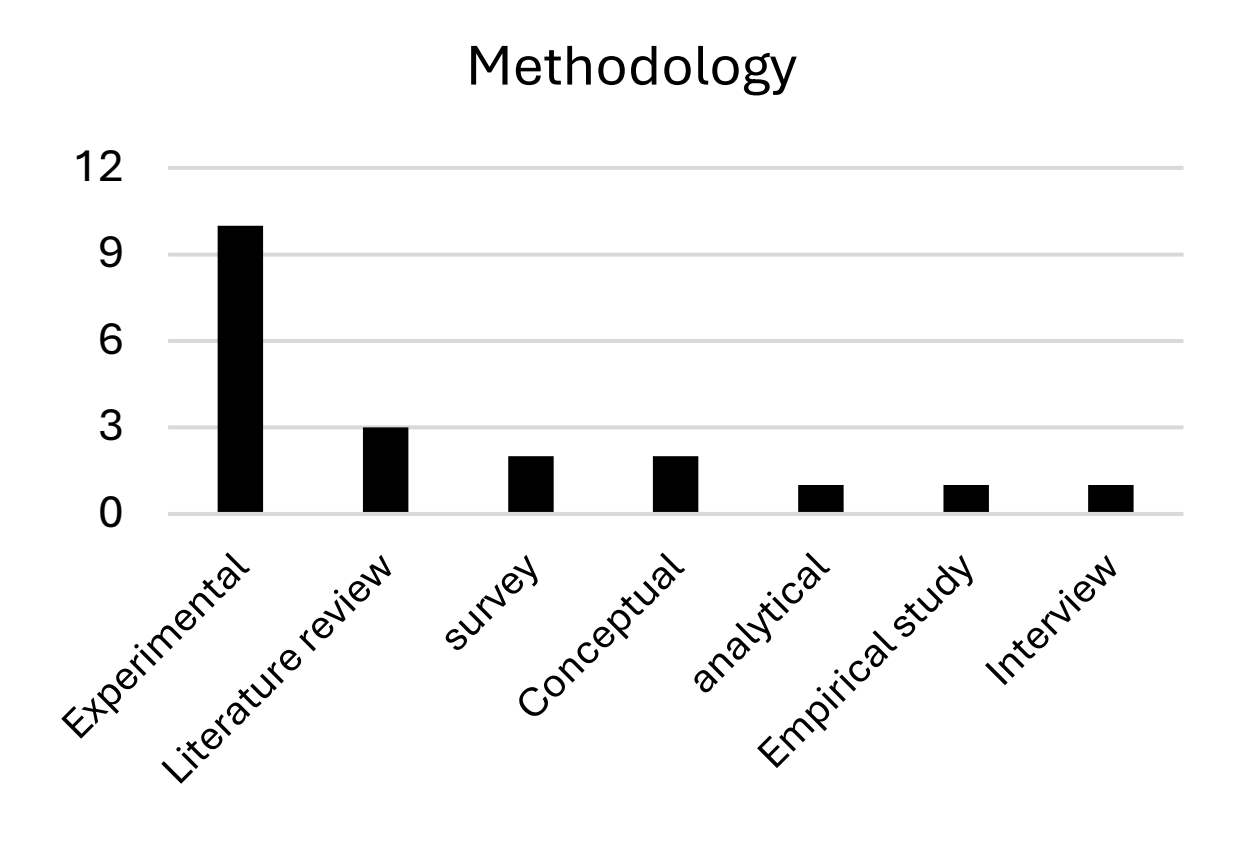

**Figure 3. A bar chart shows the methodologies used in the reviewed papers.**

## Results

Initially, we present the type of methodologies, the year of publications, and the topics covered in the selected papers. The definitions of XAI are compiled from the selected papers with an illustration of a word cloud to show the word frequencies. Lastly, we listed the challenges in the field of XAI.

This review thoroughly analysed scholarly works related to the use of XAI within the education sector. Our comprehensive review identified 19 articles focusing on XAI in education and showcasing a variety of research methods. A year-by-year breakdown of these studies is presented in Figure 2, starting with four publications in 2021, increasing to six in 2022, and nine studies until Aug 2023. Figure 3 shows the methodologies used, with experimental as the predominant one with 10 studies.

## *XAI Definitions*

By reviewing existing literature, we collated various XAI definitions to pinpoint similarities and differences. 15 distinct definitions of XAI are shown in Table 1 below:

| No | XAI Definition | Source |
|---|---|---|
| 1 | methods that produce transparent explanations and reasons for AI systems' decisions, to increase trust and promote accountability and ethics in using AI. | (Khosravi et al., 2022) |
| 2 | a field of research focused on machine learning models that can provide clear and interpretable explanations of their decisions and predictions. | (Nagy & Molontay, 2023) |
| 3 | methods and techniques in the application of AI such that human experts can understand the results of the solution. | (Farrow, 2023) |
| 4 | the ability of machines to explain their decisions and actions in a way that humans can understand. | (Cortiñas-Lorenzo & Lacey, 2023) |
| 5 | the set of methods and techniques in the application of AI such that the results of the solution can be understood by human experts. | (Fiok et al., 2022) |
| 6 | the set of tools and techniques used to make the decision-making process of machine learning models more transparent and interpretable. | (Kar et al., 2023) |
| 7 | a group of techniques and methods that aim to provide transparency and interpretability to AI models and their decision-making processes. | (Jang et al., 2022) |
| 8 | the ability of an AI system to provide clear and transparent explanations for its decision-making processes, which is crucial for building trust and accountability. | (Rachha & Seyam, 2023) |
| 9 | a movement, initiatives, and actions addressing AI transparency and trust. | (Embarak, 2022) |
| 10 | a set of techniques and methods in applying AI technology such that human experts can understand the results of the solution. | (Hanif et al., 2021) |
| 11 | a type of AI that is designed to be more transparent and understandable. | (Kobusingye et al., 2023) |
| 12 | the development of AI systems and models that can provide understandable descriptions of their decision-making strategies. | (Long et al., 2023) |
| 13 | an approach to developing AI models that humans can easily understand and interpret. | (Hasib et al., 2022) |
| 14 | a technique used to analyze the learning process of an individual student, evaluate important factors that affect learning effectiveness, and derive decision rules to predict student learning outcomes. | (Chou, 2021) |
| 15 | AI systems and algorithms that are transparent, interpretable, and able to provide explanations for their decisions and actions. | (Arnold et al., 2022) |

**Table 1. XAI Definitions**

Figure 4 displays a word cloud of the most used words in these definitions, illustrating the essence of XAI in the selected articles.

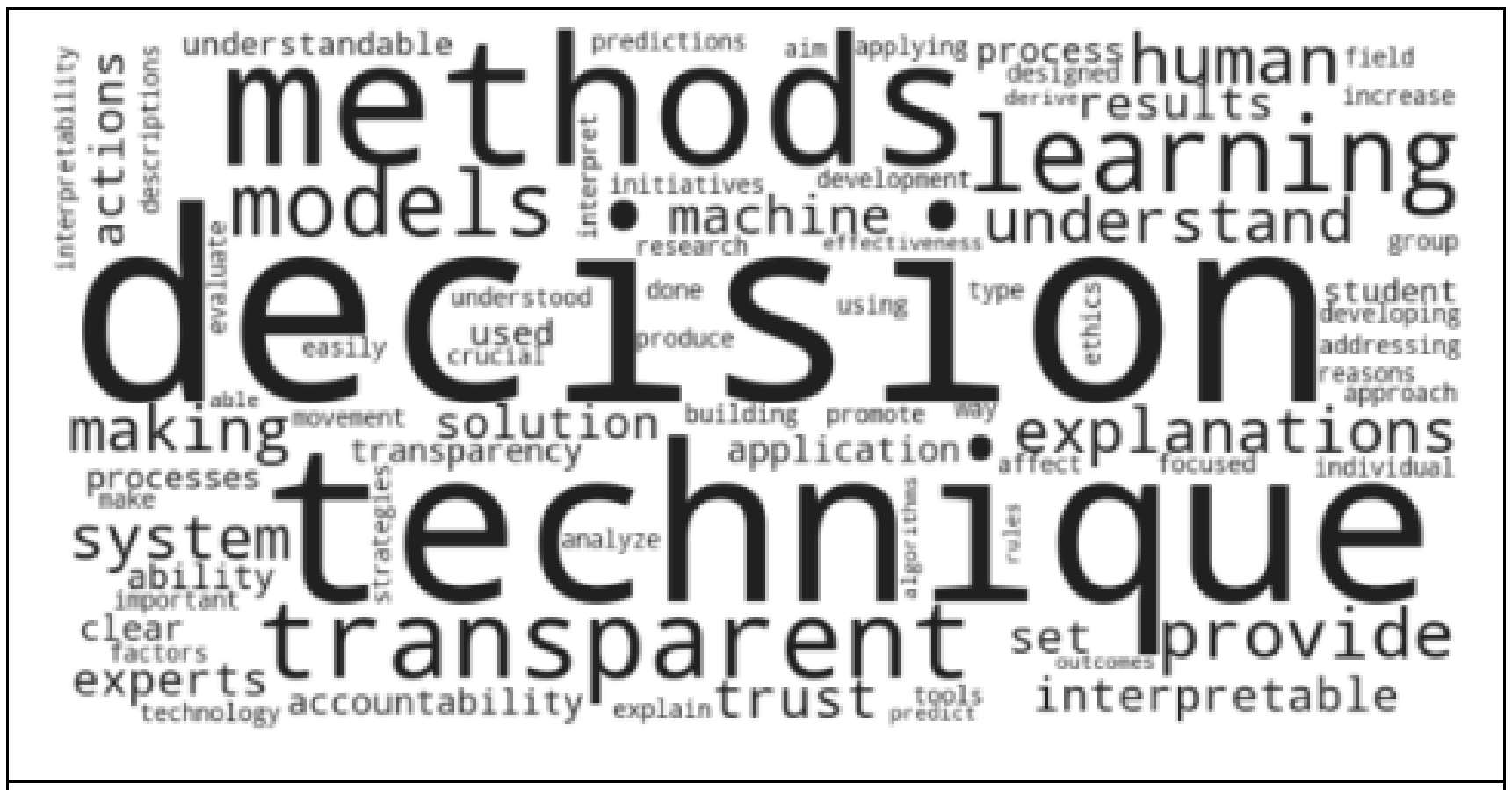

**Figure 4. Word cloud of the XAI definitions.**

## *Challenges*

In this review, a total of 62 challenges are discussed. These challenges are categorised based on the thematic analysis proposed by (Braun & Clarke, 2012), which follows six phases. We constructed a classification system based on the types of challenges identified.

The first phase involved familiarising ourselves with the relevant data by thoroughly examining selected publications to grasp their context and relevance to research objectives. Subsequently, we extracted challenges related to XAI in education from these articles and organised them into a table to discern thematic characteristics for further analysis.

In the second phase, we assigned initial codes to each challenge using a labelling system denoted as "CH#" we re-arranged them by looking at the similar meanings they portray. This coding helped us identify patterns and potential categorisation from the selected publications, with specific keywords emerging consistently across different challenges, highlighting their potential unifying themes. This step is vital for building the base of the thematic approach (Byrne, 2022).

The third phase is to create a theme that is meaningful to classify the challenges into specific categories by analysing patterns and keywords repeated across the selected papers. Identifying and grouping these keywords form cohesive themes to categorise the challenges. We labelled these repeated terminologies as "Keywords," this approach allowed us to conceptualise the data meaningfully, confirming the importance of these keywords in clustering the challenges under specific categories.

The fourth phase entailed a detailed analysis of each challenge to determine its alignment with the established themes and attribute it to one or more categories based on this thematic approach we created in the previous step. Challenges that did not fit any theme—such as those calling for further research or highlighting the costs of developing XAI systems—were classified under "Other". These represent a variety of challenges that did not consistently appear across the literature.

The fifth phase is to name the themes by grouping the keywords used to attribute the challenges into a specific category. We named the categories after reviewing multiple publications that discuss the XAI challenges and found the proper names for each category: Kamath and Liu (2021) decompose the explainability (CH1) into sub-principles centred on the nature of explanations themselves, including understandability, comprehensibility, interpretability, and transparency. Furthermore, the ethical (CH2) considerations in XAI are based on eight principles: accountability, responsibility, transparency, fidelity, bias, causality, fairness, and safety (Hanif et al., 2021).

The technical (CH3) challenges are classified based on the keywords related to the models, algorithms, and features, to name a few. These terms represent the technical aspect of the XAI, as guided by Futia and Vetrò (2020). They claimed that the technical challenges in XAI include developing approaches usable for non-experts, manipulating the recursive algorithms underlying deep learning models, and lacking insights into the internal working models of these models.

Moreover, the Human-Computer-Interaction (CH4) focuses on the stakeholders and their relationship with the systems. Khosravi et al. (2022) outlined three key objectives of HCI in education that professionals and AI developers can collaborate to achieve: developing user-centred interfaces, fostering the creation of explainable algorithms, and integrating user participation into the design process. Based on these objectives, we categorised certain challenges into HCI (CH4). While HCI plays a crucial role in a development phase, its focus on the user experience might lead to it being perceived as less technically relevant and, therefore, being isolated from core technical considerations (CH3). Next, trustworthy (CH5) concerns are classified based on accuracy, transparency, reliability, privacy, robustness, safety, security, mitigating bias, fairness, and accountability (Phillips et al., 2021).

Finally, we attribute any challenge that is associated with 'strategy', 'policy', 'standard', 'framework', etc., to the policy and guideline (CH6) category. The policies and guidelines organise and ensure the AI systems comply with all relevant requirements in relevant domains they adopt; it makes the AI-driven systems ethical, tends to provide transparent and highly accurate explanations, ensures it is developed in a safe and secure environment and is user-friendly. Table 2 shows random examples of the challenges we collected and used when applying the thematic approach.

| XAI Challenge | CH1 | CH2 | CH3 | CH4 | CH5 | CH6 | CH7 | source |
|---|---|---|---|---|---|---|---|---|
| There are no standardized methods for addressing explainability. | ✓ | | | | | ✓ | | (Hasib et al., 2022) |
| Ensuring that AI models are aligned with educational standards and ethical considerations and how to balance it benefits especially about data privacy and security. | | ✓ | ✓ | | ✓ | ✓ | | (Khosravi et al., 2022) |
| An XAI model's limitations include limited scope, generalization, complexity, computational resource requirements, and reliability. | ✓ | ✓ | ✓ | | | | | (Melo et al., 2022) |
| The potential for prediction bias or inaccuracy if the data sets used to train the AI models are biased or incomplete. | | ✓ | ✓ | | ✓ | | | (Ghosh et al., 2023) |
| The trade-off between model accuracy and interpretability could be influenced by different factors, such as the type of learning data and the complexity of the problem. | ✓ | | | | ✓ | | | (Chou, 2021) |
| There are no designing AI systems that are transparent, explainable, and accessible for non-expert users. | ✓ | ✓ | | ✓ | ✓ | | | (Long et al., 2023) |
| Identifying ... relevant set of features that are meaningful to individuals ... | | | ✓ | ✓ | | | | (Kar et al., 2023) |
| The development and deployment of AI in education can be expensive for educational institutions, impeding its widespread adoption. | | | | | | | ✓ | (Rachha & Seyam, 2023) |

**Table 2: Example of Random Sample (Challenges and the Categorisation Attribution)**

The result of this thematic approach can conclude that the challenges associated with Explainability (labelled as CH1) and ethical consideration (CH2) are most discussed, with a frequency of 23 and 20 times, respectively, in the selected papers. The key concepts within CH1 include explanatory capacity, complexity, interpretability, explanation, understandability, meaningfulness, and transparency. The CH2 involves critical issues such as bias, privacy, ethical considerations, discrimination, reliability, fairness, and accountability. Next, technical challenges (CH3) were discussed, with 19 instances that revolved around security, data quality, computational resource requirements, models, feature selection, and factor-based features. This category deals with the underlying technical aspects that support AI systems. In this sequential rank, Human-Computer Interaction (HCI), denoted as (CH4), appears 17 times concerning user interface design, stakeholders, users, socio-technical systems, and human factors. This category emphasises designing interfaces and interactions that accommodate human needs and behaviours, facilitating effective communication between AI systems and their users.

## Explainability (CH1)

Many algorithms are black-box models, and it is difficult to explain or interpret their results, especially for educational stakeholders like decision-makers and educators (Ghosh et al., 2023; Hanif et al., 2021).

One of the major challenges in the field of XAI is balancing explanatory fidelity and complexity (Wang et al., 2021). If the explanation is simplified for end-users, less information provided for this purpose may mislead the system's decision-making process. Also, the complexity may increase due to the multi-disciplinary teams involved in the development process (Kar et al., 2023), which is a challenge to balance between the complexity and the system explanations (Chou, 2021). Addressing model complexity and mitigating risks associated with misinterpretation (Fiok et al., 2022) are also critical explainability challenges.

| Code | Category | Count | Keywords | Source |
|---|---|---|---|---|
| CH1 | Explainability | 23 | Explanatory, complexity, interpretable, explanation, understandability, meaningful, transparency | (Kamath & Liu, 2021) (Wang et al., 2021) |
| CH2 | Ethical | 20 | Bias, privacy, transparency, ethical, discrimination, reliability, fairness, accountability | (Hanif et al., 2021) (Arnold et al., 2022) |
| CH3 | Technical | 19 | Security, data quality, computational resource requirements, models, Feature selection, factor-based, features | (Futia & Vetrò, 2020) (Rachha & Seyam, 2023) (Chou, 2021) |
| HC4 | HCI | 17 | User interface, stakeholders, users, socio-technical, human | (Khosravi et al., 2022) (Fiok et al., 2022) |
| CH5 | Trustworthy | 13 | trust, inequalities, inaccuracy, coherent, faithful, transparency | (Phillips et al., 2021) (Khosravi et al., 2022) |
| CH6 | Policy and Guideline | 9 | strategies, policy, standards, standardized, framework, Guidelines | (Melo et al., 2022) (Rachha & Seyam, 2023) |
| CH7 | Other | 4 | | |

**Table 3. Categorisation of Challenges Faced in XAI for Education.**

Another significant issue is the trade-off between model transparency, interpretability, and accuracy (Chou et al., 2022; Fiok et al., 2022; Hasib et al., 2022), and there is a need to find a balance among them (Kar et al., 2023). In some cases, more complex models may be necessary to achieve higher accuracy, but these models are often less interpretable (Hasib et al., 2022). Khosravi et al. (2022) claim that complexity leads to issues of transparency, accountability, and trust. Preserving key information in the explanations is important to help users understand the relationship between inputs and outputs. Developing methods that can generate meaningful and helpful explanations for users in different contexts (Cortiñas-Lorenzo & Lacey, 2023) are also included in explainability challenges

## Ethical (CH2)

As AI technology is relatively new, balancing the benefits of AI models with ethical considerations, particularly in terms of data privacy and security, presents a significant challenge (Khosravi et al., 2022). The use of AI in education raises ethical concerns, particularly regarding data privacy, bias, and algorithmic fairness (Rachha & Seyam, 2023). Ensuring the preservation of student privacy and data security while employing AI-driven systems in education is crucial (Jang et al., 2022), along with ensuring that XAI systems do not contribute to social or economic inequality and that decisions made by these systems do not violate ethical norms or harm individuals or society (Arnold et al., 2022).

Another major ethical concern is the identification and mitigation of biases in model and feature selection, which stem from hidden assumptions within AI (Wang et al., 2021). These biases might arise during the data collection process or from the foundational assumptions of the AI models themselves (Khosravi et al., 2022). Additionally, even if an AI model is explainable, biases or inaccuracies may still occur due to the datasets used in training (Ghosh et al., 2023; Hasib et al., 2022).

Ensuring that AI models adhere to educational standards and ethical norms is imperative. Nagy and Molontay (2023) highlight that privacy concerns also extend to collecting and analysing sensitive student data, which must be handled with the utmost care. They also emphasise that personalised interventions must be implemented equitably to prevent exacerbating existing inequalities.

## Technical (CH3)

Implementing AI models in the education sector requires significant technical expertise in Machine Learning (ML), often beyond the resources available within many educational institutions (Rachha & Seyam, 2023). The integration of multiple models presents challenges for educators who are unfamiliar with data mining and ML techniques, potentially necessitating additional resources and expertise (Chou, 2021). Additionally, the task of selecting and preparing appropriate learning data for model training involves domain-specific knowledge, which can be particularly challenging as it requires understanding both educational contexts and technical data handling (Hasib et al., 2022).

Maintaining data quality and accurately identifying critical variables are significant technical hurdles exacerbated by the growing volume of educational data (Khosravi et al., 2022). Furthermore, determining which factor-based models are most effective for specific educational contexts and selecting practical features to identify students at risk of underperformance accurately pose additional challenges (Jang et al., 2022). Notably, poor data quality can lead to inaccurate predictions, thus undermining the effectiveness of AI in educational settings (Rachha & Seyam, 2023).

## Human-Computer Interaction (HCI) (CH4)

Artificial intelligence in education (AIED) systems are socio-technical, merging both technical and social components, which could lead to unintended consequences (Farrow, 2023). Over-reliance on these technologies may diminish the role of human teachers, potentially leading to a less personalised and more standardised educational experience (Ghosh et al., 2023). It is crucial for educators to cautiously interpret the system's predictions, considering additional factors that might affect student performance (Jang et al., 2022).

To facilitate the effective use of XAI systems, it is imperative to develop user interfaces that are intuitive and accessible to all users, including non-technical stakeholders such as teachers, students, and administrators (Arnold et al., 2022; Khosravi et al., 2022). Creating these interfaces involves not only providing clear explanations and actionable guidance but also ensuring that they support easy and efficient interactions between users and AI systems (Fiok et al., 2022; Wang et al., 2021). Addressing these HCI challenges is essential for integrating AI effectively into educational contexts.

## Trustworthy (CH5)

Transparency and explainability, biases and discrimination, privacy and security, and the potential impact of AIED on human agency, autonomy, and responsibility (Farrow, 2023) are significant challenges in developing trustworthy AI systems. Assessing the trade-off between accuracy and transparency or explanation quality (Fiok et al., 2022) is another challenge in this domain.

The need to balance the interpretability and accuracy of the model (Kar et al., 2023) is a crucial challenge in developing trustworthy and ethical AI systems. Ethical concerns related to student privacy and data security need to be ensured (Jang et al., 2022) while ensuring that the training data used to build the model is diverse and representative, which can help improve the accuracy and bias of the AI model (Hasib et al., 2022).

## Policy and Guideline (CH6)

Challenges in policy and guideline formulation within the context of XAI in education include the lack of standardised methods to evaluate, explore, and compare multimodal datasets (Cortiñas-Lorenzo & Lacey, 2023). Additionally, comprehensive frameworks are absent to assess AI literacy and address the ethical and policy issues that emerge with AI integration (Long et al., 2023). Ensuring that AI models comply with educational standards and ethical considerations presents another significant challenge (Khosravi et al., 2022).

There is also a pressing need to establish guidelines for educational interventions that are based on online behaviours and to provide a comprehensive evaluation across various educational stakeholders, privacy and security concerns, technical challenges related to seamlessly integrating the different technologies and the legal and ethical implications (Embarak, 2022; Jang et al., 2022). The lack of uniform standards complicates the integration of AI technologies into education, as it hinders interoperability and compatibility across different educational systems (Rachha & Seyam, 2023). Specific domains, such as addressing student dropouts, Melo et al. (2022) highlight the development of effective prevention strategies as a crucial challenge. These strategies must account for the lack of specific data, policy variations, and the necessity to address multiple factors with tailored actions.

## Other Challenges (CH7)

The adoption of XAI in education faces significant hurdles, primarily due to the resources required. Developing and deploying AI technologies in educational settings is not only time-consuming but also costly. Institutions must contend with the expense of producing thousands of instance labels—a tedious and costly process—and the broader costs of AI implementation, which may hinder its widespread adoption (Ghai et al., 2021; Rachha & Seyam, 2023).

Additionally, effective use of AI requires substantial expertise in ML and AI. This expertise is often segmented; for instance, the integration of human knowledge inputs is typically isolated from other parts of ML development, requiring iterative and asynchronous coordination with data scientists who act as mediators (Ghai et al., 2021). Another significant challenge lies in tailoring educational interventions. Identifying and implementing the most effective type of intervention for individual students consistently across various educational contexts remains a complex issue (Nagy & Molontay, 2023). Moreover, as XAI is a relatively new and continually evolving field, there is a pressing need for further research and development to explore its practical applications and integration into real-world educational systems (Kar et al., 2023).

# Discussion

XAI seeks to remove the opaque workings of algorithmic models, fostering trust in their decision-making processes. However, XAI contains a multifaceted set of components, making it challenging to prioritise one over others. As shown in the Venn diagram below, the intersection between these components presents not only how they are interlinked with each other but also indicates how policies and guidelines cover all of them for a trustable AI-driven system.

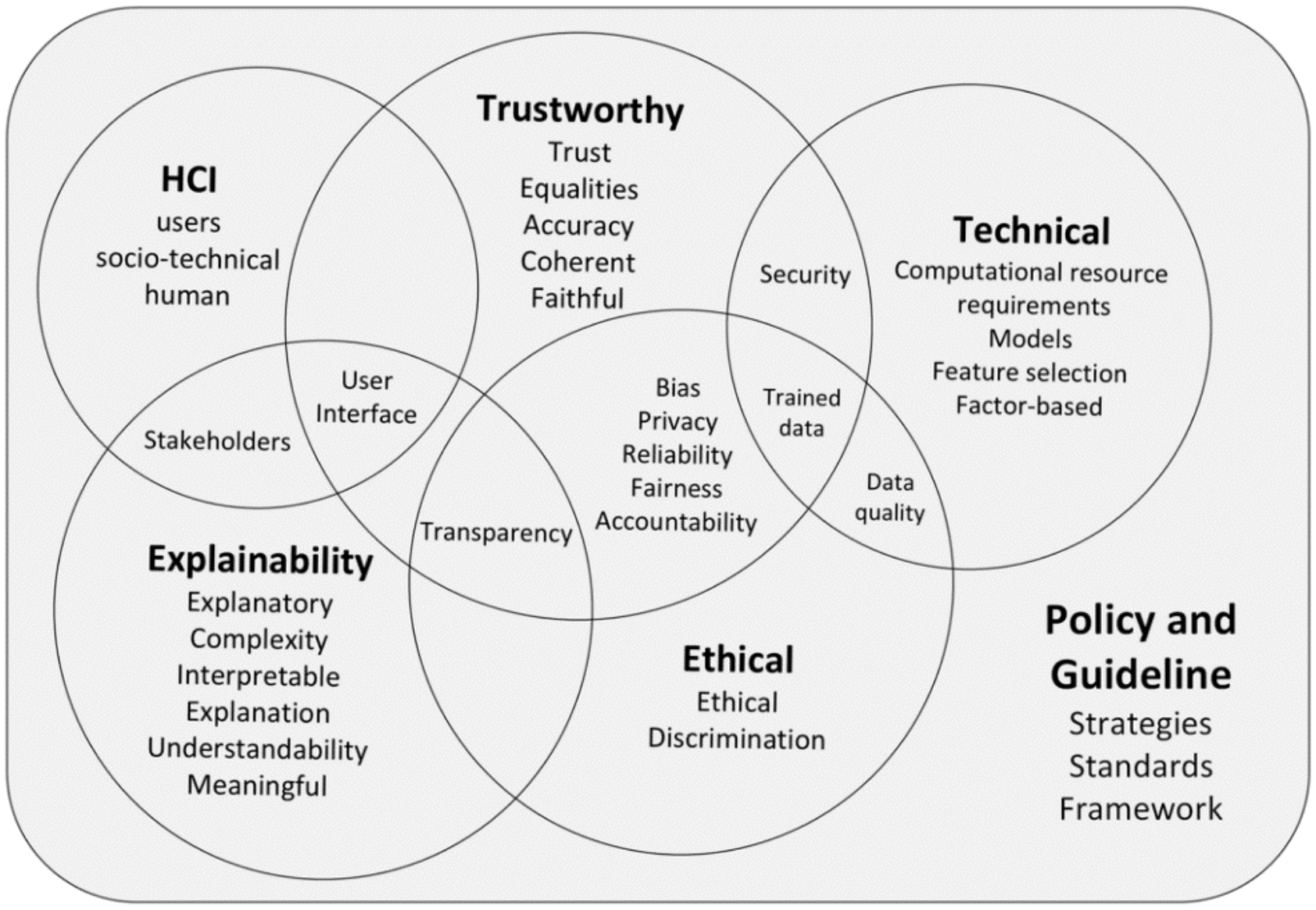

**Figure 4. The Venn diagram shows the intersection between the different XAI components.**

## *XAI definitions*

The diversity in definitions of XAI highlighted by various researchers underscores the complexity and richness of the field. For instance, some researchers focus primarily on the methods, tools and techniques that reveal the opaque AI algorithm. This perspective, as noted (Chou, 2021; Farrow, 2023; Fiok et al., 2022; Hanif et al., 2021; Jang et al., 2022; Khosravi et al., 2022), emphasises the technical aspect of making AI decisions understandable to humans. Also, Kar et al. (2023) describe XAI as a set of tools that reflect a practical approach, focusing on the tangible means through which transparency and interpretability can be achieved. This pragmatic perspective aligns well with the operational needs of developers and end-users who require tools that can be directly applied to enhance the transparency of AI systems. The word cloud in Figure 4, emphasising terms such as "decision," "technique," and "methods," visually underlines these common themes.

In contrast, another group of researchers, including Arnold et al. (2022) and Cortiñas-Lorenzo and Lacey (2023), conceptualise XAI through the lens of the AI system's inherent capabilities to produce explanations that are not just interpretable but also transparent. This suggests a shift from only understanding outputs post-hoc to designing systems that inherently operate in a comprehensible manner but understandable. Further, the definitions by Hasib et al. (2022), Kobusingye et al. (2023), and Long et al. (2023) point towards the development and design of AI systems with built-in transparency. This approach not only facilitates comprehension but also fosters trust and accountability in AI operations, which necessitates systems that individuals can trust and understand on a fundamental level.

In contrast, some definitions extend beyond the operational and technical aspects of XAI to consider its broader impacts and implications. For example, Nagy and Molontay (2023) and Embarak (2022) frame XAI as a field of research and a movement aimed at fostering ethical practices and ensuring the responsible use of AI. This broader view emphasises the societal and ethical dimensions of XAI, including governance, policy-making, and public trust. In synthesising these views, it is evident that while the core objectives of transparency and interpretability are widely recognised, the means of achieving these objectives and the emphasis on different aspects of the AI systems vary.

## *Explainability and Interpretability*

Explainability is the heart of the XAI because it is the way that it is used to achieve its goals. Explainability refers to the ability to understand and present the reasons behind an AI model's decisions or predictions in a way that is understandable to humans. Another definition is the ability to explain the entire process of a model from input to output, and it fixes the black-box issue and makes models transparent (Onose, 2023). However, if the explanations are not transparent or understandable, this might raise a problem in achieving the goal of the XAI. In addition, the inability to fully explain the performance of state-of-the-art ML algorithms highlights a knowledge gap between research and practical applications; this gap impedes the understanding of complex relationships that AI uncovers, relationships that often surpass the limits of human reasoning (Arrieta et al., 2020). The term explainability contains multiple concepts connected in order to provide valid explanations. The main concepts include but are not limited to interpretability, complexity, and transparency.

Interpretability is another factor that could affect the model's explainability. However, there is a slight difference between the meaning of explainability and interpretability. Interpretability is the ability to explain a system's results; then, we can evaluate whether these results are reasonable in light of the auxiliary criteria for the context (Doshi-Velez & Kim, 2017). This assumes that the explanations are provided in understandable terms and do not need further explanations (Guidotti et al., 2018). Interpretability also ensures AI systems are trustworthy by facilitating regulatory compliance and exposing potential biases or shortcomings that could lead to discriminatory or erroneous outputs (Arrieta et al., 2020).

Both explainability and interpretability must be utilised to deliver meaningful results that are easy for the end user to understand. If the stakeholder meaningfully comprehends the results provided by XAI, the system's trustworthiness will increase. By developing more transparent and comprehensible models, XAI techniques enable one to comprehend the reasoning behind an AI system's decision-making process (Hanif et al., 2021). Also, it aims to create models that are more transparent and understandable while still achieving a high standard of learning or performance (Farrow, 2023). Vary techniques are used to achieve the goals of XAI based on the end-user requirements. Each one has its own way of explaining the ML models.

In addition, complexity is another factor that can prevent the explanations from being delivered clearly to the audience. It is the inability to offer a complete comprehension of the AI system's general operation and decision-making process (Rachha & Seyam, 2023). Especially if the audience is ordinary people who are not experts in data science or algorithms and the explanations are a bunch of numbers and indicators that would make the explanations useless due to the audience's less experience in understanding the results.

Moreover, transparency represents the intrinsic interpretability of a model, whereas understandability quantifies the extent to which a human can comprehend a model's decision-making process (Arrieta et al., 2020). XAI provides descriptive explanations, which lead to the transparency that could be provided by the technical team (McDermid et al., 2021). Explainability, interpretability, and complexity are connected to ensure that the XAI provides the proper explanations in a convenient way that suits the target audience to achieve trustworthiness.

## *Trustworthy*

Trustworthiness is considered a requirement for applying and developing AI systems to be used by people and communities (European Commission, 2019). Trust is the conviction that everything people place their confidence in will not damage them in any way (Chamola et al., 2023). Trusted systems mean that the system achieves multiple requirements. European Commission (2019) outlined guidelines consisting of seven technical requirements and three components for trustworthy AI systems. Seven technical requirements: a) Human agency and oversight; b) Technical robustness and safety; c) Privacy and data governance; d) Transparency; e) Diversity, non-discrimination, and fairness; f) Societal and environmental well-being; and g) Accountability. Three components: a) lawful, b) ethical and c) robust.

Trustworthiness is a challenging concept that requires multiple skills to ensure the AI system obeys the regulations and follows the guides to ensure the system provides the highest degree of accuracy and accountability to the end user. It requires multiple levels of design and development, including audibility, minimisation and reporting of negative impact, trade-offs, and redress. The meaning of trustworthiness for XAI is to utilise trustworthiness to provide trustful, complete, and transparent explanations.

## *Ethical*

The ethical considerations represent a challenge to those who use and develop AI systems. Ethical AI systems are developed by prioritising fairness, accountability, transparency, privacy, accessibility, and human well-being. Also, they avoid bias and complexity with an emphasis on saving users' rights. Ethical AI refers to developing, deploying, and applying AI that ensures conformity with ethical norms, such as fundamental rights, special moral entitlements, ethical principles, and related core values for achieving a trustworthy AI (European Commission, 2019). Utilising AI in education presents ethical issues, especially regarding data privacy, bias, and algorithmic fairness (Rachha & Seyam, 2023). It can be noticed that transparency is a common concept between the ethical and explainability categories.

Bias and discrimination in AI systems are ethical issues, and they pose risks that might result in inaccurate results and lead to poor decisions. It usually results from the data on which the system is trained. Biases may arise from errors in the data collection process or from the underlying assumptions of AI models (Khosravi et al., 2022). Although the bias could have occurred unintentionally, there is a tendency to harm and damage the reputation of stakeholders (Roselli et al., 2019). In addition, without sufficient attention to ethical considerations and openness, artificial intelligence has the potential to replicate existing patterns of social inequity, technical bias, and discriminatory views (Wang et al., 2023). One example of bias is discrimination against users in some form, either colour, financial status, or ethnicity. Discrimination could lead to unfair outcomes based on biased inputs. However, XAI has the ability to tackle this issue as Hofeditz et al. (2022) demonstrate that XAI can influence discrimination in AI-driven decisions and suggest that XAI explanations moderate the effect of AI recommendations, which reduces the negative reactions to results perceived as biased.

Moreover, protecting user privacy is essential to keep the systems aligned with user safety, which keeps the end-user data safe from access and exposure by unauthorised parties. Ensuring privacy can obtain the user's trust in the AI systems. In today's world, many emerging technologies may put people at risk and breach their privacy (Mills & Bali, 2023).

Educational data are usually sensitive, and students have the right to be informed about how the AI system processes the data. When establishing effective tailored intervention educational applications, privacy considerations regarding the collection and analysis of sensitive student data must be taken into account (Nagy & Molontay, 2023).

## *Policy and Guideline*

One of the challenges in XAI is that various policies, guidelines, and frameworks are in place. The most pressing challenge is that no commonly agreed framework or methodology exists for developing and implementing XAI models. This makes it difficult to compare and assess the quality and transparency of AI outputs (Hasib et al., 2022). Also, the lack of a unified framework refers to the difficulties of evaluating the effectiveness of computing systems, particularly ones that include many modalities (Cortiñas-Lorenzo & Lacey, 2023). For XAI, the absence of a unified framework makes defining the concepts difficult, complex and disarray. Palacio et al. (2021) state that a continued, unnoticed, yet recognised challenge in this field is the absence of agreement on the terminology related to XAI.

Although many efforts from global organisations, nations, and researchers to provide guidelines and policies still vary based on the perspective and contexts for which those are formulated. Nannini et al. (2023) examine the policies and guidelines for XAI within the EU, US, and UK. In the EU, the AI Act adopts a risk-based strategy, requiring transparency and explainability, particularly in high-risk AI systems. In contrast, the US emphasises self-regulation through ethical AI guidelines that promote transparency and safeguard civil rights, with a key focus on the AI Bill of Rights. The UK, meanwhile, prioritises industry-specific guidance by establishing comprehensive frameworks for AI explainability, especially within government and public sector contexts. They found that the regulations limit tackling XAI's complexity and sociotechnical impact, which shrinks the opportunities to develop and implement in new fields. Also, despite these regulatory efforts, significant challenges remain regarding the practical implementation and standardisation of XAI practices.

Another example is where researchers propose clinical XAI guidelines for medical image analysis to apply understandability, trustworthiness, and computational efficiency (Jin et al., 2023).

In an educational setting, the policies and guidelines are essential to ensure that the technology adoption protects the stakeholder's rights and meets their expectations. However, there are demands from researchers to come up with a unified framework and guidelines in Explainable AI in Education (XAIED) to unify the terminologies and ensure that AI in education complies with transparency and accountability (Rachha & Seyam, 2023). In addition, finding the right legal framework for explaining AI decisions is challenging. Contextual organisational factors and setting a baseline between explainability, complexity, stakeholder requirements, and expertise may introduce further tensions to enhanced transparency (Nannini et al., 2023).

## *Technical*

The reviewed papers highlight technical challenges within the domain of XAI, identifying 17 specific issues. Research indicates these challenges typically stem from poor data quality, limited computational resources, and obstacles in selecting appropriate features. Technical issues can influence various facets of an AI-driven system. For example, certain problems might introduce bias or compromise the system's accuracy. Bias can occur due to flawed data collection methods or the inherent assumptions of the AI model's developers, resulting in erroneous outputs (Khosravi et al., 2022). Addressing issues of privacy, security, and data quality surrounding the collection and use of sensitive data for model training and validation (Wang et al., 2021) is a significant challenge.

Another issue has to do with computational limitations in XAI, which hinder its development and deployment. XAI models can be too specific, difficult to generalise, complex, computationally expensive, and sometimes unreliable (Melo et al., 2022). Moreover, the technical issues are significant and need to be tackled to avoid affecting other limitations. The technical challenges are not necessarily related to response time or making the system easy to use; they affect the heart of the goal of AI-driven systems. One technical challenge could impact explainability, fairness, and trustworthiness.

However, there are many efforts to mitigate these technical issues, such as developing new XAI methods that can provide more accurate results. Also, many publications discuss methods to reduce the complexity and computational time of the sophisticated ML black-box model, such as the Dimensional Reduction technique (Li et al., 2023).

## *Human-Computer Interaction (HCI)*

One key technical challenge in the field of XAI is Human-Computer Interaction (HCI). We found that HCI challenges often relate to user interface, socio-technical and stakeholder requirements. HCI refers to how society works with technology. It covers how computers affect society, their role in education, and virtually all other areas where humans and computers interact (Kieras, 1990). The main aspect of XAI in education is to identify what and how the explanations are delivered to the end user (Khosravi et al., 2022). Also, it is important to be aware that educational systems have both technical and social dimensions, and it is crucial to avoid unintended negative effects by carefully developing those systems. XAI reviewed papers outlining the need to improve the user interface of the explanations to deliver understandable results. In the context of education, there is a need to create interfaces that are easy to understand for various stakeholders such as teachers, students, and administrators (Khosravi et al., 2022).

Research indicates that the HCI community does not make much effort to contribute to the XAI research, which could be affected by the social sciences and human behaviour (Rachha & Seyam, 2023). Also, it is essential to have an idea about the end-user background and the ability to understand the outputs of the XAI techniques. The explainability and interpretability of the black-box models and algorithms have difficulty being understandable by non-technical end-users (Ghosh et al., 2023). Also, it is important to identify and select the most relevant set of features that are meaningful to individuals who are not data scientists, which represents the importance of involving all stakeholders in developing the XAI systems (Kar et al., 2023).

### *Impacts of XAI Challenges on Educational Stakeholders*

The problems concerning XAI in education affect diverse educational stakeholders (teachers, students, and administrators) that need to be understood for effective implementation. Teachers face the challenge of integrating XAI systems due to the requirement for educational and technical expertise. Additionally, the sophisticated interfaces are not intuitive thereby making it difficult to understand AI predictions that can result in misrepresented actions. Moreover, mistrust only makes it worse as teachers' reliance on AI depends upon how accurate, transparent or unbiased the predictions are.

Moreover, there are ethical issues for students regarding equity, data privacy and security, including concerns about collecting sensitive information and processing it through biased algorithms, resulting in unfair treatment. These problems are further exacerbated by unreliable policies, leading to different applications of AI tools that may be unjustifiable. Also, the XAI systems need to tackle the different student's needs based on personal, family, and educational factors.

Also, administrators struggle with deploying and maintaining AI systems that meet both education standards and ethical mandates because of the lack of a standardised framework. Policymakers should also include regulations that will ensure transparency, accountability, and fairness of such machines while balancing between model accuracy and interpretation.

### *Recent Development of Applied XAI in Education*

Although XAI is a relatively recent development, its applications in educational contexts are increasingly recognised. Many publications highlight the potential benefits of XAI in this area, presenting various case studies and applications. For example, Ghosh et al. (2023) employed XAI to gain insights into the study interests of engineering students in higher education. Similarly, Melo et al. (2022) and Nagy and Molontay (2023) used XAI to evaluate factors contributing to school dropouts. Additionally, XAI has been applied to predict student performance, as demonstrated by Jang et al. (2022) and Hasib et al. (2022), both of whom focused on predicting school student outcomes. Finally, Chou (2021) utilised XAI to enhance the assessment of learning effectiveness in online education.

However, these case studies and applications are used to show how the XAI is beneficial in the education sector; most of the real-world educational systems that utilise the XAI as a function are still in the developmental or early stages of adoption (Gade et al., 2020). Also, many frameworks and methodologies introduced by researchers are still theoretically discussed, which presents the gap between the research and the practical environment, which requires an integration between them.

## Conclusion

This study delves into the increasing importance of Explainable AI (XAI) in the educational sphere, using a systematic analysis to uncover diverse definitions and challenges associated with them. The literature reviewed identified 15 interpretations of XAI. We observed that no singular definition comprehensively encompasses all facets of XAI, with each study providing a definition reflective of its particular viewpoint or focus. However, we noticed that there are some similarities and differences between those definitions. This lack of a standardised definition for the terms associated with XAI processes leads to ambiguity, particularly as definitions related to ethics, trustworthiness, technical aspects, and explainability vary and often overlap, making the application and complete understanding of these terms challenging.

This review has discovered that XAI faces 62 challenges, which have been classified using thematic analysis into seven categories (ethical, HCI, technical, trustworthy, policy and guideline, explainability, and others) to enhance our understanding of XAI's impact on education. Those challenges must be tackled to ensure that AI systems are safely and comprehensively used. Researchers conclude that ethics in AI are frequently overlooked, although organisations are taking some action to address this issue (Capgemini Worldwide, 2019). Accessibility and equity are other limitations that need to be solved where not all users may have access to AI systems, and there is a need to guarantee that any AI models employed are visible, explainable, and devoid of bias (Alasadi & Baiz, 2023). We also found that there is a need for more research to enrich the policies, guidelines, and regulations related to XAI. Guidelines need to account for the technical specifications and skills in order to ensure the system develops with high standards and avoid the major issues that cause the system to lose its trustworthiness. Furthermore, the categories used to organise the challenges intersected and influenced each other; for example, enhanced explainability can affect technical operations, and increased transparency can have implications for ethical considerations. We also found that there is a need for further research to clarify these concepts, enabling policymakers to define the necessary criteria to ensure AI systems are in line with their objectives, regulatory requirements, and the expectations of stakeholders.

## Research Agenda

This review concentrated on identifying the definitions and challenges associated with XAI within the educational sector. Nevertheless, broadening this examination to encompass additional fields such as finance, healthcare, cybersecurity, and transportation would be beneficial. These sectors can utilise the methodology used in this review to uncover the XAI definitions and challenges that may impact regulatory compliance and trust within the different sectors. In addition, researchers can study the impact of existing and upcoming regulations on the development and deployment of XAI systems by examining how XAI can help organisations comply with regulations like GDPR. Also, to find practical insights, researchers can conduct case studies on the implementation of XAI across various industries and document the challenges, solutions, and outcomes.

Moreover, our investigation sourced publications from three databases; there is an opportunity for the research community to extend their exploration to include additional databases, potentially uncovering more insights relevant to education. Furthermore, our study was confined to the year 2023, and it is anticipated that future research will continue to evolve in the areas of XAI, which will be crucial for addressing identified challenges. Additionally, our collection of definitions for XAI comes at a time when these technologies are in their nascent stages, and it is likely that these definitions will evolve as a more precise understanding and more advanced technologies are applied to XAI.

All these demands illustrate how the current situation is an opportunity to search in and enrich it with more research and experiments, which ensure the AI systems are trustworthy and achieve the goals that are made to. The continuous discussion among researchers tackles and mitigates the challenges that XAI faces while simultaneously enhancing our grasp of XAI. The development of XAI depends on this variety of opinions since it drives ongoing research and development of techniques meant to make AI systems not only efficient but also trustworthy.

## References

Alasadi, E. A., & Baiz, C. R. (2023). Generative AI in Education and Research: Opportunities, Concerns, and Solutions. *Journal of Chemical Education*, *100*(8), 2965-2971. https://doi.org/10.1021/acs.jchemed.3c00323

Ali, S., Abuhmed, T., El-Sappagh, S., Muhammad, K., Alonso-Moral, J. M., Confalonieri, R., Guidotti, R., Del Ser, J., Díaz-Rodríguez, N., & Herrera, F. (2023). Explainable Artificial Intelligence (XAI): What we know and what is left to attain Trustworthy Artificial Intelligence. *Information Fusion*, *99*, 101805.

Arnold, O., Golchert, S., Rennert, M., & Jantke, K. P. (2022). Interactive Collaborative Learning with Explainable Artificial Intelligence. International Conference on Interactive Collaborative Learning,

Arrieta, A. B., Díaz-Rodríguez, N., Ser, J. D., Bennetot, A., Tabik, S., Barbado, A., Garcia, S., Gil-Lopez, S., Molina, D., Benjamins, R., Chatila, R., & Herrera, F. (2020). Explainable Artificial Intelligence (XAI): Concepts, taxonomies, opportunities and challenges toward responsible AI. *Information Fusion*, *58*, 82-115. https://doi.org/https://doi.org/10.1016/j.inffus.2019.12.012

Braun, V., & Clarke, V. (2006). Using thematic analysis in psychology. *Qualitative research in psychology*, *3*(2), 77-101.

Braun, V., & Clarke, V. (2012). Thematic analysis. In H. Cooper, P. M. Camic, D. L. Long, A. T. Panter, D. Rindskopf, & K. J. Sher (Eds.), *APA handbook of research methods in psychology* (Vol. 2, pp. 57-71). American Psychological Association. https://doi.org/https://doi.org/10.1037/13620-004

Byrne, D. (2022). A worked example of Braun and Clarke's approach to reflexive thematic analysis. *Quality & quantity*, *56*(3), 1391-1412.

Capgemini Worldwide. (2019). *Why addressing ethical questions in AI will benefit organizations*. https://www.capgemini.com/wp-content/uploads/2021/02/AI-in-Ethics_Web-1.pdf

Chamola, V., Hassija, V., Sulthana, A. R., Ghosh, D., Dhingra, D., & Sikdar, B. (2023). A Review of Trustworthy and Explainable Artificial Intelligence (XAI). *IEEE Access*, *11*, 78994-79015. https://doi.org/10.1109/ACCESS.2023.3294569

Chou, T. (2021). Apply explainable AI to sustain the assessment of learning effectiveness. 12th International Multi-Conference on Complexity, Informatics and Cybernetics, IMCIC 2021,

Chou, Y.-L., Lin, Y.-H., Lin, T.-Y., You, H. Y., & Chang, Y.-J. (2022, 2022). Why Did You/I Read but Not Reply? IM Users' Unresponded-to Read-Receipt Practices and Explanations of Them.*CHI '22*

Cortiñas-Lorenzo, K., & Lacey, G. (2023). Toward Explainable Affective Computing: A Review. *IEEE Transactions on Neural Networks and Learning Systems*, 1-21. https://doi.org/10.1109/TNNLS.2023.3270027

Doshi-Velez, F., & Kim, B. (2017). Towards A Rigorous Science of Interpretable Machine Learning. In: arXiv.

Embarak, O. H. (2022). Internet of Behaviour (IoB)-based AI models for personalized smart education systems. *Procedia Computer Science*, *203*, 103-110. https://doi.org/https://doi.org/10.1016/j.procs.2022.07.015

European Commission. (2019). *ETHICS GUIDELINES FOR TRUSTWORTHY AI: High-Level Expert Group on Artificial Intelligence*. https://ec.europa.eu/futurium/en/ai-alliance-consultation.1.html

Farrow, R. (2023). The possibilities and limits of XAI in education: a socio-technical perspective. *Learning, Media and Technology*, 1-14.

Fiok, K., Farahani, F. V., Karwowski, W., & Ahram, T. (2022). Explainable artificial intelligence for education and training. *The Journal of Defense Modeling and Simulation*, *19*(2), 133-144.

Futia, G., & Vetrò, A. (2020). On the Integration of Knowledge Graphs into Deep Learning Models for a More Comprehensible AI—Three Challenges for Future Research. *Information*, *11*(2), 122. https://www.mdpi.com/2078-2489/11/2/122

Gade, K., Geyik, S., Kenthapadi, K., Mithal, V., & Taly, A. (2020). Explainable AI in industry: Practical challenges and lessons learned. Companion Proceedings of the Web Conference 2020,

Galhotra, S., Pradhan, R., & Salimi, B. (2021). Explaining black-box algorithms using probabilistic contrastive counterfactuals. Proceedings of the 2021 International Conference on Management of Data,

Ghai, B., Liao, Q. V., Zhang, Y., Bellamy, R., & Mueller, K. (2021). Explainable Active Learning (XAL): Toward AI Explanations as Interfaces for Machine Teachers. *Proc. ACM Hum.-Comput. Interact.*, *4*(CSCW3), 1-28. https://doi.org/10.1145/3432934

Ghosh, S., Kamal, M. S., Chowdhury, L., Neogi, B., Dey, N., & Sherratt, R. S. (2023). Explainable AI to understand study interest of engineering students. *Education and Information Technologies*, 1-16.

Guidotti, R., Monreale, A., Ruggieri, S., Turini, F., Pedreschi, D., & Giannotti, F. (2018). A Survey Of Methods For Explaining Black Box Models. In (pp. 1-42): arXiv.

Hanif, A., Zhang, X., & Wood, S. (2021, 2021/10//). A Survey on Explainable Artificial Intelligence Techniques and Challenges. 2021 IEEE 25th International Enterprise Distributed Object Computing Workshop (EDOCW),

Hasib, K. M., Rahman, F., Hasnat, R., & Alam, M. G. R. (2022, 2022/01/26/). A Machine Learning and Explainable AI Approach for Predicting Secondary School Student Performance. 2022 IEEE 12th Annual Computing and Communication Workshop and Conference (CCWC),

Hofeditz, L., Clausen, S., Rieß, A., Mirbabaie, M., & Stieglitz, S. (2022). Applying XAI to an AI-based system for candidate management to mitigate bias and discrimination in hiring. *Electronic Markets*, *32*(4), 2207-2233. https://doi.org/10.1007/s12525-022-00600-9

Jang, Y., Choi, S., Jung, H., & Kim, H. (2022). Practical early prediction of students' performance using machine learning and eXplainable AI. *Education and Information Technologies*, *27*(9), 12855-12889.

Jin, W., Li, X., Fatehi, M., & Hamarneh, G. (2023). Guidelines and evaluation of clinical explainable AI in medical image analysis. *Medical Image Analysis*, *84*, 102684. https://doi.org/https://doi.org/10.1016/j.media.2022.102684

Kamath, U., & Liu, J. (2021). *Explainable artificial intelligence: an introduction to interpretable machine learning* (1 ed., Vol. 2). Springer Cham. https://doi.org/https://doi.org/10.1007/978-3-030-83356-5

Kar, S. P., Das, A. K., Chatterjee, R., & Mandal, J. K. (2023). Assessment of learning parameters for students' adaptability in online education using machine learning and explainable AI. *Education and Information Technologies*, 1-16.

Kaur, D., Uslu, S., Rittichier, K. J., & Durresi, A. (2022). Trustworthy artificial intelligence: a review. *ACM computing surveys (CSUR)*, *55*(2), 1-38.

Khosravi, H., Shum, S. B., Chen, G., Conati, C., Tsai, Y.-S., Kay, J., Knight, S., Martinez-Maldonado, R., Sadiq, S., & Gašević, D. (2022). Explainable Artificial Intelligence in education. *Computers and Education: Artificial Intelligence*, *3*, 100074. https://doi.org/10.1016/j.caeai.2022.100074

Kieras, D. E. (1990). An Overview of Human-Computer Interaction. *Journal of the Washington Academy of Sciences*, *80*(2), 39-70. http://www.jstor.org/stable/24531047

Kobusingye, B. M., Dorothy, A., Nakatumba-Nabende, J., & Marvin, G. (2023). Explainable Machine Translation for Intelligent E-Learning of Social Studies. 2023 7th International Conference on Trends in Electronics and Informatics (ICOEI),

Li, X., Chen, D., Xu, W., Chen, H., Li, J., & Mo, F. (2023). Explainable dimensionality reduction (XDR) to unbox AI 'black box'models: A study of AI perspectives on the ethnic styles of village dwellings. *Humanities and Social Sciences Communications*, *10*(1), 1-13.

Long, D., Roberts, J., Magerko, B., Holstein, K., DiPaola, D., & Martin, F. (2023). AI Literacy: Finding Common Threads between Education, Design, Policy, and Explainability. Extended Abstracts of the 2023 CHI Conference on Human Factors in Computing Systems,

McDermid, J. A., Jia, Y., Porter, Z., & Habli, I. (2021). Artificial intelligence explainability: the technical and ethical dimensions. *Philosophical Transactions of the Royal Society A: Mathematical, Physical and Engineering Sciences*, *379*(2207), 20200363. https://doi.org/10.1098/rsta.2020.0363

Melo, E., Silva, I., Costa, D. G., Viegas, C. M., & Barros, T. M. (2022). On the use of explainable artificial intelligence to evaluate school dropout. *Education Sciences*, *12*(12), 845. https://doi.org/https://doi.org/10.3390/educsci12120845

Mills, A., & Bali. (2023). How do we respond to generative AI in education? Open educational practices give us a framework for an ongoing process. *Journal of Applied Learning & Teaching*, *6*(1). https://doi.org/10.37074/jalt.2023.6.1.34

Nagy, M., & Molontay, R. (2023). Interpretable Dropout Prediction: Towards XAI-Based Personalized Intervention. *International Journal of Artificial Intelligence in Education*, 1-27. https://doi.org/10.1007/s40593-023-00331-8

Nannini, L., Balayn, A., & Smith, A. L. (2023). Explainability in AI Policies: A Critical Review of Communications, Reports, Regulations, and Standards in the EU, US, and UK. Proceedings of the 2023 ACM Conference on Fairness, Accountability, and Transparency,

Onose, E. (2023). Explainability and Auditability in ML: Definitions, Techniques, and Tools. *MLOps Blog*. https://neptune.ai/blog/explainability-auditability-ml-definitions-techniques-tools#:~:text=Explainability%20in%20machine%20learning%20means,applies%20to%20all%20artificial%20intelligence.

Page, M. J., McKenzie, J. E., Bossuyt, P. M., Boutron, I., Hoffmann, T. C., Mulrow, C. D., Shamseer, L., Tetzlaff, J. M., Akl, E. A., & Brennan, S. E. (2021). The PRISMA 2020 statement: an updated guideline for reporting systematic reviews. *Bmj*, *372*.

Palacio, S., Lucieri, A., Munir, M., Ahmed, S., Hees, J., & Dengel, A. (2021). Xai handbook: towards a unified framework for explainable AI. Proceedings of the IEEE/CVF International Conference on Computer Vision,

Phillips, P. J., Phillips, P. J., Hahn, C. A., Fontana, P. C., Yates, A. N., Greene, K., Broniatowski, D. A., & Przybocki, M. A. (2021). *Four principles of explainable artificial intelligence* (NIST IR 8312). https://nvlpubs.nist.gov/nistpubs/ir/2021/NIST.IR.8312.pdf

Rachha, A., & Seyam, M. (2023, 2023/04/01/). Explainable AI In Education : Current Trends, Challenges, And Opportunities. SoutheastCon 2023,

Rai, A. (2020). Explainable AI: From black box to glass box. *Journal of the Academy of Marketing Science*, *48*, 137-141.

Roselli, D., Matthews, J., & Talagala, N. (2019, 2019/05/13/). Managing Bias in AI. WWW '19: The Web Conference,

Wang, C., Wang, K., Bian, A. Y., Islam, R., Keya, K. N., Foulds, J., & Pan, S. (2023). When Biased Humans Meet Debiased AI: A Case Study in College Major Recommendation. *ACM Trans. Interact. Intell. Syst.*, *13*(3). https://doi.org/10.1145/3611313

Wang, P. Y., Galhotra, S., Pradhan, R., & Salimi, B. (2021). Demonstration of generating explanations for black-box algorithms using Lewis. *Proceedings of the VLDB Endowment*, *14*(12), 2787-2790.